%% file: prd_xib0_v3.tex
\documentclass[aps, prd, preprint, amsmath, amssymb,showpacs]{revtex4}

\usepackage{epsfig}
\usepackage{graphicx}
\usepackage{dcolumn}
\usepackage{bm}
\usepackage{setspace}

\begin{document}


\renewcommand{\thetable}{\Roman{table}}

\begin{flushright}
Fermilab-Pub-11-323-E-PPD \\
\end{flushright}

\title{
Observation of the $\Xi_b^0$ Baryon}

\input{March2010_Authors.tex}

\begin{abstract}
The observation of  the  bottom, strange baryon $\Xi^0_b$ through the
decay chain  $\Xi^0_b \rightarrow \Xi_c^+ \, \pi^-$, where $\Xi_c^+ \rightarrow
\Xi^- \, \pi^+ \, \pi^+$,  $\Xi^- \rightarrow \Lambda \, \pi^-$, and $\Lambda
\rightarrow p \, \pi^-$, is reported 
using data corresponding to an integrated luminosity 
of 4.2 fb$^{-1}$ from $p\bar p$ collisions
at $\sqrt{s}=1.96$ TeV  recorded with the Collider Detector at Fermilab. A
signal of $25.3^{+5.6}_{-5.4}$ candidates  is observed whose probability of
arising  from a background fluctuation is $3.6 \times 10^{-12}$,   corresponding
to 6.8 Gaussian standard deviations. The $\Xi^0_b$ mass is measured to be 
$5787.8\pm5.0(\textrm{stat})\pm1.3(\textrm{syst})$ MeV/$c^2$. In addition, the
$\Xi^-_b$ baryon  is observed through the process  $\Xi^-_b \rightarrow \Xi_c^0
\, \pi^-$, where $\Xi_c^0 \rightarrow \Xi^-\, \pi^+$,  $\Xi^- \rightarrow
\Lambda \, \pi^-$, and $\Lambda \rightarrow p \, \pi^-$.
\end{abstract}
\maketitle

The quark model has had great success in describing
the spectroscopy of hadrons.  
For the $c$ and $b$ mesons, all of the 
ground states have been observed \cite{PDG}.
The spectroscopy of $c$ baryons also agrees well with the quark model, 
and a  rich
spectrum of baryons containing $b$ quarks is predicted \cite{Jenkins}.
Until recently, direct observation of $b$ baryons 
has been limited to a single state, the
$\Lambda^0_b$ (quark content $|udb \, \rangle$) \cite {PDG}.  
The accumulation of large data sets from the Tevatron has improved
this situation and made possible the observation of the
$\Xi^-_b$ ($|dsb \, \rangle$) \cite{D0_Xi_b, CDF_Xi_b},
the $\Sigma^{(*)}_b$ states ($|uub \, \rangle, |ddb \, \rangle$)
\cite{CDF_Sigma_b},
and the $\Omega_b$ ($|ssb \, \rangle$)\cite{D0_Omega_b,CDF_Omega_b}.

In this paper, we report the observation of
an additional heavy baryon 
and the measurement of its mass.
The decay properties of this state are consistent with the weak decay
of a $b$ baryon.  We interpret the result as the observation
of the $\Xi_b^0$ baryon ($|usb \, \rangle$). 
This measurement is made in
$p\overline{p}$ collisions 
at a center of mass energy of 1.96 TeV using the
Collider Detector at Fermilab (CDF II), by fully reconstructing 
the decay chain 
$\Xi_{b}^0 \rightarrow$ $\Xi_c^+ \, \pi^-$, where
$\Xi_c^+ \rightarrow \Xi^- \, \pi^+\, \pi^+$, 
$\Xi^- \rightarrow \Lambda \, \pi^-$,
and $\Lambda \rightarrow p \, \pi^-$.
Charge conjugate modes are included implicitly.
In addition, we observe 
the $\Xi_b^-$ through the  similar decay chain
$\Xi^-_{b} \rightarrow$ $\Xi_c^0 \, \pi^-$, where
$\Xi_c^0 \rightarrow \Xi^- \, \pi^+$, $\Xi^- \rightarrow \Lambda \, \pi^-$,
and $\Lambda \rightarrow p \, \pi^-$.
These studies use
a data sample corresponding to an integrated luminosity of
4.2 fb$^{-1}$ and constitute the first exclusive reconstruction of the
$\Xi_b^0$ and the first for the $\Xi_b^-$ in this decay channel.  


The CDF II detector has been described in detail elsewhere 
\cite{CDF_detector}.  
This analysis relies upon the tracking system  
that operates inside a 1.4 T solenoidal magnetic field.
A five-layer silicon detector (SVX\ II) 
measures track positions at
radii of 2.5 to 10.6 cm to provide high precision impact parameter 
measurements. 
Each of these layers
provides a transverse measurement
and a stereo measurement of $90^{\circ}$
(three layers) or  $\pm 1.2^{\circ}$ (two layers)
with respect to the beam direction.
An open-cell drift chamber (COT) 
covers the radial region from
43 cm to 132 cm and provides track momentum meaurement.

Data acquisition is triggered by a system designed to collect
particle candidates 
that decay with lifetimes characteristic of heavy flavor hadrons.
The first level of the trigger system requires two
tracks in the COT with transverse momentum $p_T > 2.0 $ GeV/$c$.
In the second level of the trigger, the silicon vertex trigger 
\cite{SVT} is used to associate SVX II data with the tracks 
found in in the COT and provides precise impact parameter resolution
(typically 40 $\mu$m) for
these tracks.  The silicon vertex trigger requires two tracks with  
impact parameters in the range 0.1-1.0 mm with 
respect to the beam and a point of intersection that is measured
with at least a 200 $\mu$m displacement transverse to the beam.


This analysis combines the trajectories of charged particles to
infer the presence of several different hadrons in the decay chains.
The decay point for each weak decay process in the decay chain is 
reconstructed and used to identify the corresponding hadron. 
Consequently, it is useful to define two quantities
that are used frequently throughout the analysis that relate
the paths of weakly decaying objects to their points of origin.
Both quantities are defined in the transverse view and
make use of the point of closest approach $\vec {r}_{c}$ 
of the particle trajectory to a point of origin $\vec {r}_{o}$ and 
of the measured
particle decay position  $\vec {r}_{d}$.
The first quantity used here is transverse flight distance $f(h)$
of hadron $h$, which is the distance
a particle has traveled in the transverse view.  
For neutral particles, flight distance is given by
$f(h) \equiv (\vec {r}_{d} - \vec {r}_{o})
 \cdot \vec{p}_{T}(h)/|\vec{p}_{T}(h)|$, where 
$\vec{p}_{T}(h)$ is the transverse momentum of the hadron candidate. 
For charged particles, the flight distance is calculated as the arc 
length in the transverse view from $\vec {r}_{c}$ to $\vec {r}_{d}$.
Flight distance is used to calculate the proper decay time
of weakly decaying states, where the decay time is given by
$t \equiv f(h) M(h)/(c \, |\vec{p}_{T}(h)|)$, 
where $M(h)$ is the reconstructed mass. 

A complementary quantity used in this analysis is transverse 
impact distance $d(h)$,
which is the distance of the point of closest approach to the 
point of origin.  For neutral particles, transverse impact distance is given by
$d(h) \equiv  |(\vec {r}_{d} - \vec {r}_{o})
\times \vec{p}_{T}(h)|/|\vec{p}_{T}(h)|$.
The impact distance of charged particles is simply the distance from
$\vec {r}_{c}$ to the point of origin.
 

The reconstruction of  $\Lambda$
candidates uses all tracks with $p_T \, >$ 0.4 GeV/$c$
found in the COT.  
Pairs of oppositely charged tracks are combined to identify these
neutral decay candidates, and silicon detector information
is not used due to the large transverse displacement of the $\Lambda$
decay. 
Candidate selection is 
based upon the mass calculated for each track pair,
which has a resolution of 1.5-2.0 MeV/$c^2$ and
is required to fall within 9 MeV/$c^2$ of the nominal $\Lambda$ mass \cite{PDG}
after the appropriate mass assignment for each track.
The proton (pion) mass is assigned 
to the track with the higher (lower) momentum.
This mass assignment is 
always correct for the $\Lambda$ candidates 
used in this analysis
because of the
kinematics of $\Lambda$ decay and the lower limit
in the transverse momentum
acceptance of the tracking system. 
Background to the  
$\Lambda$ ($c\tau=7.9$ cm) \cite{PDG} is reduced by requiring
the transverse flight distance 
of the  $\Lambda$ from the beam position to be
greater than 1.0 cm, which corresponds to typically $0.6 \, \sigma_f$, where 
$\sigma_f$ is the flight distance resolution.


For events that contain a $\Lambda$ candidate,
the remaining tracks reconstructed in the COT, again without additional
silicon information,
are assigned the pion mass, and
$\Lambda \, \pi^-$  combinations 
are identified that are consistent with the decay process
$\Xi^- \rightarrow \Lambda \, \pi^-$.  
Several features of the track topology are used to reduce the
background to this process.
In order to obtain the best possible 
 mass resolution for $\Xi^-$ candidates, 
the reconstruction requires a convergent  
fit of the three tracks that simultaneously 
constrains the $\Lambda$
decay products to the $\Lambda$ mass and the
$\Lambda$ trajectory to intersect with the helix of 
the $\pi^-$ originating from the $\Xi^-$ candidate.
The $\Lambda \, \pi^-$ mass obtained from this fit
has a resolution comparable to the $\Lambda$ and is required to 
fall within 9 MeV/$c^2$ of the nominal $\Xi^-$ mass \cite{PDG}.
In addition,
the flight distance of the $\Lambda$ candidate with respect to the
reconstructed decay point of the $\Xi^-$ candidate
is required to exceed 1.0 cm.
Similarly, due to the long lifetime of the 
weakly decaying $\Xi^-$ ($c\tau$ = 4.9 cm)
 \cite{PDG}, a transverse flight distance
of at least 1.0 cm (which typically corresponds to $1.0 \, \sigma_f$) 
with respect to the beam position is required.  

In some instances, the intersection of the $\pi^-$ helix 
with the $\Lambda$ trajectory produces a situation
where two
$\Lambda \, \pi^-$ 
vertices satisfy the constrained fit and displacement
requirements.
In addition, the complexity of the $\Xi^-$ and $\Lambda$ decays allows for
occasional combinations where the proper identity of the three tracks
is ambiguous.  
A single, preferred candidate is chosen by retaining only
the fit combination with the highest probability of satisfying
the constrained fit.

The kinematics of hyperon decay and the lower $p_T$ limit of 0.4 GeV/$c$
on the decay daughter tracks force the majority of $\Xi^-$ 
candidates to have $p_T >$ 1.5 GeV/$c$.  
This fact, along with the long lifetime of the $\Xi^-$,
results in a significant fraction of the hyperon candidates having
decay points located several centimeters radially outward from the 
beam position.
Therefore, we are able to
refine the $\Xi^-$  reconstruction by making use of the 
improved determination of the trajectory that can be obtained by tracking 
these particles in
the silicon detector.  The $\Xi^-$ candidates have an additional
fit performed with the three tracks  that simultaneously 
constrains
both the $\Lambda$ and $\Xi^-$  masses of the 
appropriate track combinations
and provides the best possible estimate of the hyperon momentum and decay
position.
The result of this fit is used to define a helix that
serves as the seed for an algorithm that associates silicon detector
hits with the $\Xi^-$ track.  
Candidates with track measurements in at least one layer of the 
silicon detector
have excellent impact distance resolution (typically 60 $\mu$m). 

\begin{figure}[p]
\psfig{figure=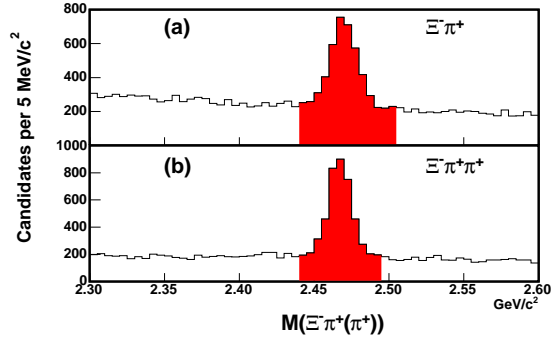, width=3.0in}
\caption{(a) The $\Xi^- \, \pi^+$  and 
(b) the $\Xi^- \, \pi^+ \, \pi^+$ mass distributions.  
The mass ranges used
for the $\Xi_c^0$ and $\Xi_c^+$ samples are indicated by the shaded areas.
\label{fig:prd_xic} }
\end{figure}

The samples of $\Xi_c^0$ and $\Xi_c^+$ candidates used in this analysis
are obtained by combining the $\Xi^-$ candidates 
that have SVX II information with additional $\pi^+$
candidates.  The $\pi^+$ candidates are tracks that have been reconstructed 
with data from at least three SVX II layers.  The $\pi^+$ used for the $\Xi_c^0$
reconstruction is required to be consistent with the trigger requirements.  
The $\Xi_c^+$
candidates are required to have at least one $\pi^+$ track consistent 
with the trigger requirements.
All $\Xi^- \, \pi^+(\pi^+)$ combinations are required to satisfy a 
constrained fit that includes mass constraints on the $\Lambda$ and $\Xi^-$
candidates.  The mass distributions of the combinations that also
satisfy $ct >$ 100 $\mu$m and $p_T > 4.0$ GeV/$c$
requirements are shown in Fig. \ref{fig:prd_xic}.
Candidates with a reconstructed mass within 30(25) MeV/$c^2$ 
of the nomimal $\Xi_c^{0}(\Xi_c^{+})$  mass are used for 
$b$ baryon reconstruction.


The $\Xi_b^{(-,0)}$ candidates are reconstructed by combining the
$\Xi_c^{(0,+)}$ candidates with $\pi^-$ candidates 
that satisfy the trigger requirements.
The $\Xi_b$ 
 candidates are required 
to have $p_T \, > $ 6.0 GeV/$c$,  restricting the sample to candidates
that are within the kinematic range where our acceptance is well modeled
\cite{CDF_Omega_b}.
All $\Xi_c \, \pi^-$ combinations are required to satisfy a 
constrained fit that includes mass constraints on the $\Lambda$, $\Xi^-$,
and $\Xi_c$ candidates.  
Combinations that are inconsistent with having
originated from the collision are rejected by imposing an upper limit on the
impact distance $d_{PV}$ of the $\Xi_b$ 
candidate measured with respect to the
primary vertex.  
In addition, the full reconstruction of the $\Xi_b$ decay chain provides
an opportunity to 
impose a requirement on the decay time of the $\Xi_c$ candidate since 
both its point of creation and decay are reconstructed.

The mean life of the charm baryons varies over a wide range and 
is large compared to the typical decay time resolution of 
20 - $60 \, \mu$m/$c$ that we measure.  
Therefore, we have chosen a selection on the $\Xi_c$ decay time that
uses the decay time resolution $\sigma_{t}$ calculated for each candidate
and the mean life of the decaying state.  The selection is developed by using
$\Lambda_b^0$ as a reference signal.
A sample of 
$\Lambda_b^0 \rightarrow \Lambda_c^+ \, \pi^-$ candidates \cite{CDF_LbLife} 
is used to optimize selection criteria for $\Lambda_c^+$ decay time
based on the mean life of the $\Lambda_c^+$ and its decay time resolution.
As a result of this study, we require that the measured decay time
of the $\Xi_c$ candidate falls within the range 
$-2\sigma_{t} \, < t \, < 3\tau + 2\sigma_{t}$ where 
$\tau$ is the mean life of the 
$\Xi_c^0(c\tau = 33 \, \mu$m) and $\Xi_c^+(c\tau = 132 \, \mu$m)
candidates. 
This requirement is found to be approximately 95\% efficient on our 
$\Lambda_b^0$ ($c\tau = 60 \, \mu$m) sample
and to reduce the background substantially.  



\begin{figure}[p]
\psfig{figure=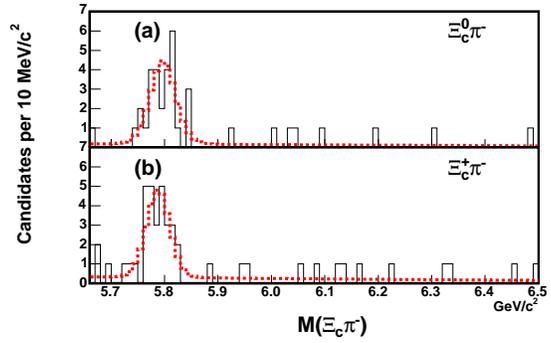, width=3.0in}
\caption{(a) The $\Xi_c^0 \, \pi^-$  and
(b) the $\Xi_c^+ \, \pi^-$ mass distributions.  
A projection of the likelihood fit is overlaid as a dashed line.
\label{fig:prd_xib} }
\end{figure}

The $\Xi_c^0 \, \pi^-$ and $\Xi_c^+ \, \pi^-$ mass distributions
with $d_{PV} \, < \,100  \, \mu$m and
$ct \, > \, 100 \, \mu$m
are shown in Fig.\ \ref{fig:prd_xib}.
These distributions 
show clear evidence of an excess near a mass of 5.8 GeV/$c^2$
with a width consistent with our expected mass measurement resolution.
The mass, yield, and significance of the $\Xi_b^{(-,0)}$ signals
are obtained by performing 
an unbinned likelihood fit on the mass distribution of candidates.
The likelihood function that is maximized has the form
${\cal L}  =  \prod_i^N \left ( 
f_sG(m_i,m_0,s_m\sigma^m_i) + (1-f_s)(a_0 + a_1 m_i)\right ),$
where $N$ is the number of candidates in the sample,
$G(m_i,m_0,s_m\sigma^m_i)$ is a Gaussian distribution with average
$m_0$ and characteristic width $s_m\sigma^m_i$ to describe the signal,
$m_i$ is the mass obtained for a single $\Xi_c^{(0,+)} \pi^-$ candidate, 
$\sigma^m_i$ is the calculated uncertainty on $m_i$,
and the $a_n$ terms model the background.  
The quantities
obtained from the fitting procedure include 
the fraction $f_s$ of the candidates identified as signal, 
the best average mass value $m_0$, 
a scale factor on the mass resolution $s_m$ to allow for inaccuracy of the
resolution estimate, 
and the values of $a_0$ and $a_1$.

For this data sample,
several variations of the fit were used to test the significance.
The first of these fits corresponds to the null signal hypothesis, and fixes
$f_s=0.0$, $s_m = 1.0$, and $m_0$ to the nominal mass of the $\Xi_b^-$.
Additional applications allow $f_s$ to float, retain the
constraints on $s_m$, and fix $m_0$ to values within 5 MeV/$c^2$ of
the nominal mass of the $\Xi_b^-$.
The value of $-2\ln{\cal L}$ for the null hypothesis
exceeds the values for the fits with variable $f_s$ by at least 
48.2 units for the 
$\Xi_b^-$ candidate sample and by 48.3 units for the
$\Xi_b^0$ candidate sample.
We interpret these as 
equivalent to a $\chi^2$ with one degree of freedom
whose 
probability of occurrence is
$3.9\times10^{-12}$ and $3.6\times10^{-12}$, corresponding to a 
significance that exceeds $6.8\sigma$ 
 for both the $\Xi_b^-$ and $\Xi_b^0$.  
We therefore interpret these results as observations of 
the processes 
$\Xi^-_b \rightarrow \Xi^0_c \, \pi^-$ and 
$\Xi^0_b \rightarrow \Xi^+_c \, \pi^-$.

Masses are obtained from the unbinned
likelihood fit with the mass and resolution parameters allowed to vary.
In addition, the mass fit was used on the $\Xi^- \, \pi^+$ and
$\Xi^- \, \pi^+\, \pi^+$ to obtain mass measurements for the
$\Xi_c^0$ and $\Xi_c^+$, which are seen to be consistent with the
nominal values \cite{PDG}.
The results of these fits 
are listed in Table\ \ref{table:mass}.

\begin{table*}[ht]
\begin{center}
\caption{Fit results obtained for $c$ and $b$ baryons.
\label{table:mass}}
\begin{tabular}{cccc}
\hline \hline
Resonance  & Yield &Mass  (MeV/$c^2$) & Resolution Scale\\
\hline
 $\Xi^0_c$       &    $2110\pm70$  &  $2470.4\pm0.3$ & $1.16\pm0.04$  \\
 $\Xi^+_c$       &    $3048\pm67$  &  $2467.3\pm0.2$ & $1.24\pm0.03$ \\
 $\Xi^-_b$       &    $25.8^{+5.5}_{-5.2}$  &  $5796.7\pm5.1$ & $1.3\pm0.2$ \\
 $\Xi^0_b$       &    $25.3^{+5.6}_{-5.4}$  &  $5787.8\pm5.0$ & $1.2\pm0.2$ \\
\hline \hline
\end{tabular}
\end{center}
\end{table*}

The accuracy of our mass measurement scale is established by 
our measurements of the $J/\psi$, $\psi(2S)$, and $\Upsilon$ masses.
These calibration points imply an accuracy of 0.5 MeV/$c^2$ on
the mass measurements of the $\Xi_b^-$ and $\Xi_b^0$.
Our fitting technique finds that our estimate of the mass resolution
on each candidate is low, as listed in Table \ref{table:mass}.  
Fits where this scale factor was fixed at 1.0 or 1.4 introduced 
shifts in our $\Xi^0_b$ mass result by as much as 1.0 MeV/$c^2$.
A fit with a fixed 20 MeV/$c^2$ Gaussian width, as implied by the
simulation, introduced a shift of only 0.2 MeV/$c^2$.
These effects are added in quadrature with the larger of the
asymmetric nominal $\Xi_c^{(0,+)}$ mass uncertainties \cite{PDG}
to yield systematic uncertainties
of 1.4 GeV/$c^2$ for the $\Xi_b^-$ and 1.3 GeV/$c^2$ for the $\Xi_b^0$
mass measurements. 

The momentum scale  uncertainty is common
to all of our mass measurements, and can be dropped as a systematic 
uncertainty of a measurement of the mass difference between
the  $\Xi_b^-$ and $\Xi_b^0$.  Our best $\Xi_b^-$ mass measurement of 
$5790.9\pm2.6(\textrm{stat})\pm0.8(\textrm{syst})$ MeV/$c^2$
\cite{CDF_Omega_b} is obtained from the $J/\psi \, \Xi^-$ final state and
has a systematic uncertainty that would be reduced to 0.6 MeV/$c^2$
without this effect.  Therefore, we measure the mass difference
$M(\Xi_b^-) - M(\Xi_b^0) = 3.1\pm 5.6(\textrm{stat})\pm1.3(\textrm{syst})$
MeV/$c^2$, where the statistical and systematic uncertainties 
of the individual measurements have been
added in quadrature.


In conclusion, we have analyzed data collected with 
the CDF II detector at the Tevatron to
observe the bottom, strange baryon $\Xi^0_b$.
The reconstruction technique
is used on the $\Xi^-_b$ as well, and the observation 
of this state provides a cross check for the analysis.
A signal of $25.3^{+5.6}_{-5.4}$ $\Xi^0_b$ candidates,  
with a significance greater than $6\sigma$, is seen in the decay channel
$\Xi^0_b \rightarrow \Xi_c^+ \, \pi^-$ 
where $\Xi_c^+ \rightarrow \Xi^- \, \pi^+ \, \pi^+$, 
$\Xi^- \rightarrow \Lambda \, \pi^-$,
and $\Lambda \rightarrow p \, \pi^-$.
The mass of this baryon
is measured to be $5787.8\pm5.0 (\textrm{stat}) \pm1.3 (\textrm{syst})$
MeV/$c^2$,
which is consistent with theoretical expectations 
\cite{Jenkins}.
In addition, we observe $25.8^{+5.5}_{-5.2}$ candidates in the process
$\Xi^-_b \rightarrow \Xi_c^0 \, \pi^-$ 
where $\Xi_c^0 \rightarrow \Xi^- \, \pi^+$.
The mass measured for the $\Xi^-_b$ is
$5796.7\pm5.1 (\textrm{stat}) \pm1.4 (\textrm{syst})$
MeV/$c^2$, which is consistent with our earlier result \cite {CDF_Omega_b}
but does not improve upon it.
Neither of these decay channels has been reported previously, and the
reconstruction of the $\Xi^0_b$ is the first observation of this baryon in any
channel.

We thank the Fermilab staff and the technical staffs of the participating
 institutions for their vital contributions. This work was supported by 
the U.S. Department of Energy and National Science Foundation; the Italian
 Istituto Nazionale di Fisica Nucleare; the Ministry of Education, Culture,
 Sports, Science and Technology of Japan; the Natural Sciences and Engineering
 Research Council of Canada; the National Science Council of the Republic of
 China; the Swiss National Science Foundation; the A.P. Sloan Foundation; 
the Bundesministerium f\"ur Bildung und Forschung, Germany; the Korean 
World Class University Program, the National Research Foundation of Korea; 
the Science and Technology Facilities Council and the Royal Society, UK; 
the Institut National de Physique Nucleaire et Physique des Particules/CNRS; 
the Russian Foundation for Basic Research; the Ministerio de Ciencia e
 Innovaci\'{o}n, and Programa Consolider-Ingenio 2010, Spain; 
the Slovak R\&D Agency; the Academy of Finland; 
and the Australian Research Council (ARC).

\end{document}

%% file: March2010_Authors.tex
\affiliation{Institute of Physics, Academia Sinica, Taipei, Taiwan 11529, Republic of China} 
\affiliation{Argonne National Laboratory, Argonne, Illinois 60439, USA} 
\affiliation{University of Athens, 157 71 Athens, Greece} 
\affiliation{Institut de Fisica d'Altes Energies, ICREA, Universitat Autonoma de Barcelona, E-08193, Bellaterra (Barcelona), Spain} 
\affiliation{Baylor University, Waco, Texas 76798, USA} 
\affiliation{Istituto Nazionale di Fisica Nucleare Bologna, $^{aa}$University of Bologna, I-40127 Bologna, Italy} 
\affiliation{University of California, Davis, Davis, California 95616, USA} 
\affiliation{University of California, Los Angeles, Los Angeles, California 90024, USA} 
\affiliation{Instituto de Fisica de Cantabria, CSIC-University of Cantabria, 39005 Santander, Spain} 
\affiliation{Carnegie Mellon University, Pittsburgh, Pennsylvania 15213, USA} 
\affiliation{Enrico Fermi Institute, University of Chicago, Chicago, Illinois 60637, USA}
\affiliation{Comenius University, 842 48 Bratislava, Slovakia; Institute of Experimental Physics, 040 01 Kosice, Slovakia} 
\affiliation{Joint Institute for Nuclear Research, RU-141980 Dubna, Russia} 
\affiliation{Duke University, Durham, North Carolina 27708, USA} 
\affiliation{Fermi National Accelerator Laboratory, Batavia, Illinois 60510, USA} 
\affiliation{University of Florida, Gainesville, Florida 32611, USA} 
\affiliation{Laboratori Nazionali di Frascati, Istituto Nazionale di Fisica Nucleare, I-00044 Frascati, Italy} 
\affiliation{University of Geneva, CH-1211 Geneva 4, Switzerland} 
\affiliation{Glasgow University, Glasgow G12 8QQ, United Kingdom} 
\affiliation{Harvard University, Cambridge, Massachusetts 02138, USA} 
\affiliation{Division of High Energy Physics, Department of Physics, University of Helsinki and Helsinki Institute of Physics, FIN-00014, Helsinki, Finland} 
\affiliation{University of Illinois, Urbana, Illinois 61801, USA} 
\affiliation{The Johns Hopkins University, Baltimore, Maryland 21218, USA} 
\affiliation{Institut f\"{u}r Experimentelle Kernphysik, Karlsruhe Institute of Technology, D-76131 Karlsruhe, Germany} 
\affiliation{Center for High Energy Physics: Kyungpook National University, Daegu 702-701, Korea; Seoul National University, Seoul 151-742, Korea; Sungkyunkwan University, Suwon 440-746, Korea; Korea Institute of Science and Technology Information, Daejeon 305-806, Korea; Chonnam National University, Gwangju 500-757, Korea; Chonbuk National University, Jeonju 561-756, Korea} 
\affiliation{Ernest Orlando Lawrence Berkeley National Laboratory, Berkeley, California 94720, USA} 
\affiliation{University of Liverpool, Liverpool L69 7ZE, United Kingdom} 
\affiliation{University College London, London WC1E 6BT, United Kingdom} 
\affiliation{Centro de Investigaciones Energeticas Medioambientales y Tecnologicas, E-28040 Madrid, Spain} 
\affiliation{Massachusetts Institute of Technology, Cambridge, Massachusetts 02139, USA} 
\affiliation{Institute of Particle Physics: McGill University, Montr\'{e}al, Qu\'{e}bec, Canada H3A~2T8; Simon Fraser University, Burnaby, British Columbia, Canada V5A~1S6; University of Toronto, Toronto, Ontario, Canada M5S~1A7; and TRIUMF, Vancouver, British Columbia, Canada V6T~2A3} 
\affiliation{University of Michigan, Ann Arbor, Michigan 48109, USA} 
\affiliation{Michigan State University, East Lansing, Michigan 48824, USA}
\affiliation{Institution for Theoretical and Experimental Physics, ITEP, Moscow 117259, Russia}
\affiliation{University of New Mexico, Albuquerque, New Mexico 87131, USA} 
\affiliation{Northwestern University, Evanston, Illinois 60208, USA} 
\affiliation{The Ohio State University, Columbus, Ohio 43210, USA} 
\affiliation{Okayama University, Okayama 700-8530, Japan} 
\affiliation{Osaka City University, Osaka 588, Japan} 
\affiliation{University of Oxford, Oxford OX1 3RH, United Kingdom} 
\affiliation{Istituto Nazionale di Fisica Nucleare, Sezione di Padova-Trento, $^{bb}$University of Padova, I-35131 Padova, Italy} 
\affiliation{LPNHE, Universite Pierre et Marie Curie/IN2P3-CNRS, UMR7585, Paris, F-75252 France} 
\affiliation{University of Pennsylvania, Philadelphia, Pennsylvania 19104, USA}
\affiliation{Istituto Nazionale di Fisica Nucleare Pisa, $^{cc}$University of Pisa, $^{dd}$University of Siena and $^{ee}$Scuola Normale Superiore, I-56127 Pisa, Italy} 
\affiliation{University of Pittsburgh, Pittsburgh, Pennsylvania 15260, USA} 
\affiliation{Purdue University, West Lafayette, Indiana 47907, USA} 
\affiliation{University of Rochester, Rochester, New York 14627, USA} 
\affiliation{The Rockefeller University, New York, New York 10065, USA} 
\affiliation{Istituto Nazionale di Fisica Nucleare, Sezione di Roma 1, $^{ff}$Sapienza Universit\`{a} di Roma, I-00185 Roma, Italy} 

\affiliation{Rutgers University, Piscataway, New Jersey 08855, USA} 
\affiliation{Texas A\&M University, College Station, Texas 77843, USA} 
\affiliation{Istituto Nazionale di Fisica Nucleare Trieste/Udine, I-34100 Trieste, $^{gg}$University of Udine, I-33100 Udine, Italy} 
\affiliation{University of Tsukuba, Tsukuba, Ibaraki 305, Japan} 
\affiliation{Tufts University, Medford, Massachusetts 02155, USA} 
\affiliation{University of Virginia, Charlottesville, Virginia 22906, USA}
\affiliation{Waseda University, Tokyo 169, Japan} 
\affiliation{Wayne State University, Detroit, Michigan 48201, USA} 
\affiliation{University of Wisconsin, Madison, Wisconsin 53706, USA} 
\affiliation{Yale University, New Haven, Connecticut 06520, USA} 
\author{T.~Aaltonen}
\affiliation{Division of High Energy Physics, Department of Physics, University of Helsinki and Helsinki Institute of Physics, FIN-00014, Helsinki, Finland}
\author{B.~\'{A}lvarez~Gonz\'{a}lez$^w$}
\affiliation{Instituto de Fisica de Cantabria, CSIC-University of Cantabria, 39005 Santander, Spain}
\author{S.~Amerio}
\affiliation{Istituto Nazionale di Fisica Nucleare, Sezione di Padova-Trento, $^{bb}$University of Padova, I-35131 Padova, Italy} 

\author{D.~Amidei}
\affiliation{University of Michigan, Ann Arbor, Michigan 48109, USA}
\author{A.~Anastassov}
\affiliation{Northwestern University, Evanston, Illinois 60208, USA}
\author{A.~Annovi}
\affiliation{Laboratori Nazionali di Frascati, Istituto Nazionale di Fisica Nucleare, I-00044 Frascati, Italy}
\author{J.~Antos}
\affiliation{Comenius University, 842 48 Bratislava, Slovakia; Institute of Experimental Physics, 040 01 Kosice, Slovakia}
\author{G.~Apollinari}
\affiliation{Fermi National Accelerator Laboratory, Batavia, Illinois 60510, USA}
\author{J.A.~Appel}
\affiliation{Fermi National Accelerator Laboratory, Batavia, Illinois 60510, USA}
\author{A.~Apresyan}
\affiliation{Purdue University, West Lafayette, Indiana 47907, USA}
% ...
\collaboration{CDF Collaboration\footnote{With visitors from $^a$Istituto Nazionale di Fisica Nucleare, Sezione di Cagliari, 09042 Monserrato (Cagliari), Italy,
$^b$University of CA Irvine, Irvine, CA  92697, USA,
$^c$University of CA Santa Barbara, Santa Barbara, CA 93106, USA,
$^d$University of CA Santa Cruz, Santa Cruz, CA  95064, USA,
$^e$CERN,CH-1211 Geneva, Switzerland,
$^f$Cornell University, Ithaca, NY  14853, USA, 
$^g$University of Cyprus, Nicosia CY-1678, Cyprus, 
$^h$Office of Science, U.S. Department of Energy, Washington, DC 20585, USA,
$^i$University College Dublin, Dublin 4, Ireland,
$^j$University of Fukui, Fukui City, Fukui Prefecture, Japan 910-0017,
$^k$Universidad Iberoamericana, Mexico D.F., Mexico,
$^l$Iowa State University, Ames, IA  50011, USA,
$^m$University of Iowa, Iowa City, IA  52242, USA,
$^n$Kinki University, Higashi-Osaka City, Japan 577-8502,
$^o$Kansas State University, Manhattan, KS 66506, USA,
$^p$University of Manchester, Manchester M13 9PL, United Kingdom,
$^q$Queen Mary, University of London, London, E1 4NS, United Kingdom,
$^r$University of Melbourne, Victoria 3010, Australia,
$^s$Muons, Inc., Batavia, IL 60510, USA,
$^t$Nagasaki Institute of Applied Science, Nagasaki, Japan, 
$^u$National Research Nuclear University, Moscow, Russia,
$^v$University of Notre Dame, Notre Dame, IN 46556, USA,
$^w$Universidad de Oviedo, E-33007 Oviedo, Spain, 
$^x$Texas Tech University, Lubbock, TX  79609, USA,
$^y$Universidad Tecnica Federico Santa Maria, 110v Valparaiso, Chile,
$^z$Yarmouk University, Irbid 211-63, Jordan,
$^{hh}$On leave from J.~Stefan Institute, Ljubljana, Slovenia, 
}}
\noaffiliation